\documentclass{aa}  

\usepackage{graphicx,natbib}

\makeatletter
\renewcommand*\aa@pageof{, page \thepage{} of \pageref*{LastPage}}
\makeatother

\usepackage{hyperref}
 \hypersetup{pdfstartview=FitH,  citecolor=blue, linkcolor=red, urlcolor=red, colorlinks=true}
\begin{document}

   \title{Fine wave dynamics in umbral flash sources}

      \author{R. Sych
          \inst{1,2}
          \and
          M. Wang\inst{1}
                                        }

\offprints{R.~Sych, \email{sych@iszf.irk.ru}}

   \institute{
        \inst{1}Yunnan Observatories, Chinese Academy of Sciences, Kunming, Yunnan 650011, China\\
                          \inst{2}Institute of Solar-Terrestrial Physics SB RAS, Irkutsk 664033, Russia\\ 
                                                \email{sych@iszf.irk.ru}}

   \date{Revised March 24, 2018; Accepted April 23, 2018}

% \abstract{}{}{}{}{} 
% 5 {} token are mandatory
 
  \abstract
   {Umbral flashes (UFs) are most common phenomenon of wave processes in sunspots. Studying the relationship between  wave time dynamics and the origin of UFs  requires further investigation of their fine spatial and height structure.}
   {We investigated the association between a short time increase in the variations of three-minute extreme ultraviolet (EUV) emission at footpoints of coronal magnetic loops and the UF emergence in sunspot umbra. The spatial structure of magnetic field lines and their inclination determine the cut-off frequency and visibility of sunspot waves.}
   {We applied the pixelized wavelet filtering (PWF) technique to analyse a cube of the images obtained by SDO/AIA at 1600 \AA, 304 \AA, and 171 \AA ~to study the spatio-temporal oscillation power distribution in individual magnetic loops. Time-distance plots were used to obtain the wave front propagation velocity.}
   {For the first time, we obtained the 2D images of the fine wave processes in magnetic structures of different spatial scales related to the UFs in sunspot.  We revealed two types of the UFs as background and local. The background UFs associated with random increasing of separate parts of wave fronts. These UFs are seen as  weak and diffuse details that ride the wave fronts without stable shapes and localization in space. The local UFs sources mainly localize near to the footpoint of magnetic loops, anchored in an umbra, along which the propagation of three-minute waves was observed. The time dynamics of flash evolution shows an increase in three-minute oscillations before the UFs peak value within one low-frequency wave train. It is shown that the maxima of three-minute oscillation trains coincide with the peak intensity of UFs. During the UFs evolution, a fine wave spatial and temporal dynamics in UFs local sources was observed.}
   {The sunspot three-minute wave dynamics showed a relation between the localization of oscillations power peak at the coronal magnetic loop footpoints and the UFs origin. The spatial structure of the UFs sources, their power, and lifetime are determined by the cut-off frequency of the waves for the detected waveguides. We concluded that UFs are a global process of short-time increase of the wave activity.}
               
                        \keywords{Sun: oscillations--Sun: UV radiation--sunspot}
  
   \maketitle
%
%-------------------------------------------------------------------

\section{Introduction}

   Sunspot oscillations are a significant phenomenon observed in the solar atmosphere. Studying the oscillations started in 1969 \citep{1969SoPh....7..351B}, when non-stationary brightenings in the CaII and K were discovered. These brightenings were termed umbral flashes (UFs). Furthermore, \citep{1972ApJ...178L..85Z} and \citep{1972SoPh...27...71G}, using observations in the $H\alpha$ line wing, discovered ring structures in sunspots. Those structures propagated from the umbral centre to the penumbral outer boundary with a three-minute periodicity. The authors referred to these background structures as running penumbral waves (RPWs). Below, at the photosphere level, the oscillation spectrum shows a wide range of frequencies with  a peak near five-minute oscillations. These frequencies are coherent, which indicates at the umbral brightness variations within this range as a whole \citep{2004A&A...424..671K}. Also, there exist low-frequency 10-40 minute components in sunspots \citep{2009A&A...505..791S, 2008ASPC..383..279B, 2013A&A...554A.146K}. Their nature has been in doubt so far.

  Observations in \cite{2002A&A...387L..13D} showed that the observed emission in magnetic loops anchored in a sunspot has an $\sim$ 172 sec frequency periodicity, which indicates that photospheric oscillations in the form of waves can penetrate through the transition zone upwards into the corona. According to \cite{1977A&A....55..239B}, the low-frequency waves oscillated at the subphotospheric level (p-mode) propagate through natural waveguides as a concentration of magnetic elements (e.g. sunspots and pores). Their oscillation period may be modified by a mechanism for the  cut-off frequency. In \cite{1984A&A...133..333Z} showed that the oscillations with a frequency lower than the cut-off frequency fade quickly. The main factor affecting the cut-off frequency is the inclination of the field lines, along which the wave propagation occurs. We can observe five-minute oscillations both in the chromosphere spicules \citep{2004Natur.430..536D}, and in the coronal loops of active regions \citep{2005ApJ...624L..61D, 2009ApJ...702L.168D}. Further investigations of low-frequency oscillations in the solar atmosphere higher layers \citep{2009ASPC..415...28W, 2009ApJ...697.1674M, 2011SoPh..272..101Y} corroborated the assumption that their emergence at such heights is a consequence of wave channelling in the inclined magnetic fields. The observed rate of the disturbance  indicates on propagation of slow magneto-acoustic waves \citep{2009A&A...505..791S, 2012SoPh..279..427K}.
        
        For high-frequency oscillations, the sources with less than three-minute period are localized in the umbra, and they decrease in size as the period decreases \citep{2008SoPh..248..395S, 2014A&A...569A..72S, 2014A&A...561A..19Y, 2012ApJ...757..160J}. Here in the umbral central part, where the field is almost perpendicular to the Sun surface and there is no field line beam divergence, we see the footpoints of the elementary magnetic loops in the form of oscillating cells \citep{2014AstL...40..576Z}. The main mechanism that determines their power is related to the presence of the subphotospheric and chromospheric resonator in the sunspot. Outside the central part, where the field inclination starts to manifest itself, the mechanism for a cut-off frequency change begins.
        
   Sunspot oscillations are also expressed in the form of UFs \citep{1969SoPh....7..351B, 1969SoPh....7..366W}, whose emission manifests itself most definitely in the kernel of chromospheric lines. A number of papers \citep{2007PASJ...59S.631N, 2007A&A...463.1153T, 2003A&A...403..277R, 2001ApJ...552..871L,  2000Sci...288.1396S, 1983SoPh...87....7T, 1981A&A...102..147K} have studied this phenomenon. \cite{2010ApJ...722..888B} assumed that UFs are induced by magneto-acoustic waves propagating upwards that are converted into shocks. Photospheric oscillations become more abrupt as the waves move into a medium with lower density and transform into a shock front, thus heating the ambient medium. The temperature in the UF source surroundings surpasses the ambient values by 1000 K, which results in brightening of individual umbral sites of the order of several arcsec. On these scales, one also observes sunspot umbral magnetic field variations, although there is no visible confirmation of field line inclination variations or their common configuration throughout these processes  \citep{2003A&A...403..277R}. The observations taken recently have shown the presence of very small-size jet-like spatial details of less than 0.1 Mm in the sunspot umbra. Their positions are apparently related to the footpoints of single magnetic loops, along which sunspot oscillations propagate \citep{2014ApJ...787...58Y}.
        
        Umbral flashes are also related to the running wave phenomenon in a sunspot penumbra. This phenomenon is observed in $H\alpha$ and He lines \citep{2007ApJ...671.1005B} and in  CaII \citep{2013A&A...556A.115D} in the form of travelling spatial structures moving horizontally, radially from the umbra towards the penumbral outer boundary \citep{2000A&A...355..375T, 2003A&A...403..277R}. The waves that propagate along field lines are non-stationary with changes in the oscillation power both in time and in space \citep{2010SoPh..266..349S}. These results in noticeable periodic emission modulation by propagating three-minute waves at the footpoints of magnetic loops. A possible response of such a modulation is the emergence of both low-frequency wave trains, and individual oscillation maxima brightnesses as  UFs.
        
        In this study, we analysed the association between the sunspot UFs source spatial distribution and the spatial structure of the field lines anchored in the umbra. To better understand the association between oscillation activation and flash emergence, we studied the dynamics of the three-minute oscillations in UFs sources. For the spatial localization of the propagating wave fronts to magnetic waveguides, we used the method of pixelized wavelet filtration (PWF technique) \citep{2008SoPh..248..395S}. The paper is arranged as follows: in Section 1, we introduce the paper subject; in Section 2, we provide the observational data and processing methods; in Section 3, we describe the data analysis and obtained results; in Section 4, we discuss the processes of the flash evolutions; and in Section 5, we make conclusions concerning the obtained results.
        
%--------------------------------------------------------------------
\section{Observations and data processing}

 To study the connection between UFs and sunspot oscillations we used the data observations of the Solar Dynamic Observatory (SDO/AIA)      \citep{2012SoPh..275...17L} obtained with a high spatial and temporal resolution. We studied the four active regions with developed sunspots at the maximum of their wave activity. To obtain the spatial location of the UFs sources in space and height we used the observations of January 26, 2015 (NOAA 12268, 01:00-04:00 UT), January 10, 2016 (NOAA 12480, 01:00-04:00 UT), and March 27, 2016 (NOAA 12526, 01:00-04:00 UT). A more comprehensive analysis was carried out for the observations of December 08, 2010 (NOAA 11131, 00:00-03:20 UT).

We used calibrated and centred images of the Sun (Lev 1.5) for various wavelengths. The observations were performed within extreme ultraviolet (EUV; 1600 \AA) and UV (304 \AA, 171 \AA) ranges with cadences of 24 sec and 12 sec, respectively. The pixel size was            0.6 \arcsec. The differential rotation of the investigated regions during the observation was removed by using the Solar Software.

We built time-distance plots along the detected UF sources to search for a correlation between the background wavefront propagation process and the UF emergence. The precise value of the revealed oscillation periods was obtained through the Fourier method. For 2D wave processing and obtaining of their time dynamics, we used the PWF technique. The spectral algorithm applied in this method enabled us to search for waves throughout the sunspot and to trace the direction of their propagation.
        
Using the helioseismologic method to calculate the time lag of propagating three-minute wavefronts relative to each other \citep{2014A&A...569A..72S} enabled us to establish the height affixment of the SDO/AIA temperature channels. The channel of the extreme ultraviolet range 1600 \AA ~records the emission at the levels of the upper photosphere and transition region with temperatures 6000 K and 100000 K, respectively. However, the main sensitivity of the channel and, correspondingly, the minimum wave lag at the upwards propagation, falls at the emission arriving from the lower atmosphere. This channel often shows dotted, fine-structure details, brightening the field line magnetic footpoint regions. The regions with a high concentration of field lines appear black, particularly near to sunspots and active regions. The 304 \AA ~(He II) channel shows  bright regions at the level of the upper chromosphere and lower transition region, where plasma has a high density. The characteristic temperature of the channel is about 50000 K. This channel is best suited to study various oscillation processes in the solar atmosphere, particularly in sunspots where the power of three-minute oscillations reaches maximum. To observe the coronal magnetic structures, we used observations with a 171 \AA ~(Fe IX) wavelength. The emission arrives from the quiet corona and from the upper transition region with the temperature of about 1000000K.

\section{Results}

We investigated the emergence of umbral short-time recurrent flashes of brightness by using the unique observational possibility of the SDO/AIA temperature channels to receive emission from different heights of the sunspot atmosphere. This allowed us to obtain, for the first time, the information on the UF source distribution throughout an umbra and to understand their height location. To test the stability of the recurrent UF source location and their seeing at different heights, we built variation maps for the SDO/AIA different temperature channels. These maps show the distribution of the signal variation relative to its mean value in each point of images throughout the observational time.

\subsection{Spatial and heights location of UFs}

\begin{figure}%[htpb]
\begin{center}
\includegraphics[width=9.0 cm]{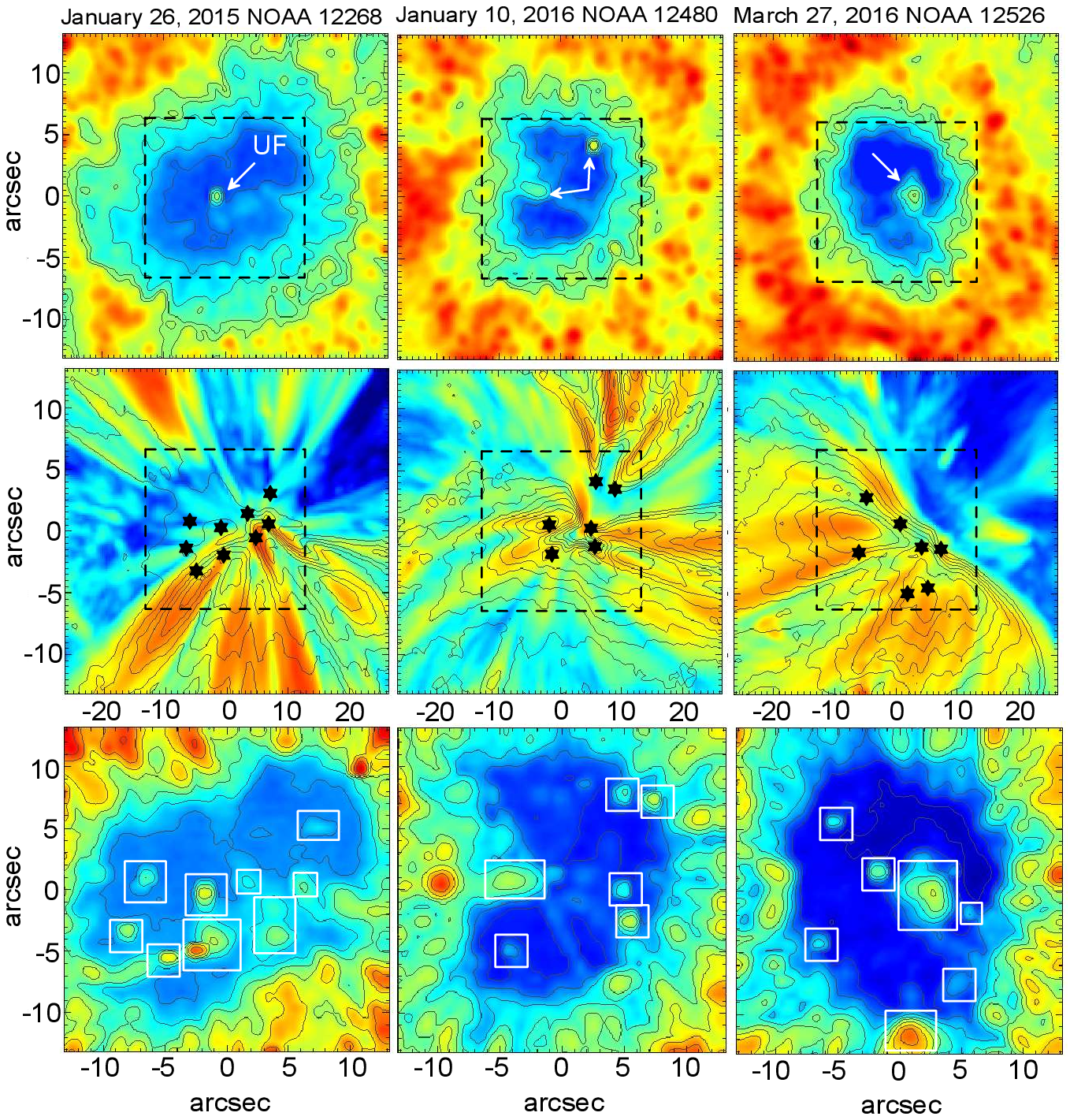}
\end{center}
\caption{Upper panels: Snapshots of the UFs in sunspot active regions on January 26, 2015 (01:57:54 UT), January 10, 2016 (01:33:52.6 UT), and March 27, 2016 (01:49:28.6 UT) obtained by SDO/AIA (1600 \AA). The broken black rectangles show the umbral regions. The arrows indicate the UFs sources. Middle panels: The corresponding sunspot regions at 171 \AA. The original maps (contours) overlapped on variation maps (colour background) of  UV emission obtained during the observation. Asterisks denote the localization of the UFs sources. Bottom panels: Scaled variation maps of the umbral regions at 1600 \AA. The small white rectangles show sources of UFs.}
\label{1}
\end{figure}

Figure~\ref{1} presents a series of sunspots images and their variation maps during the emergence of separate bright UFs obtained by SDO/AIA at 1600 \AA ~and 171 \AA. The observational time was about three hours for each day during the four days of observation. The number of the obtained images for one day was 450 frames at a 24-sec temporal resolution. Similar images were also obtained in the 304 \AA ~and 171 \AA ~channels, where the temporal resolution was 12 seconds. The number of frames was 900. This observational material is adequate to compile with confidence the statistical material both by UF number and by the location in the umbral area. The umbral regions are shown by the dash squares. To increase the visibility of the umbral weak brightening sources, we used the log scale. This enabled us to record weak processes of the umbral background wavefront propagation and to study their association with the UF emergence. This procedure was applied to all studied SDO/AIA temperature channels. This allowed us to obtain time cubes of images and to make the films, in which the dynamics of the umbral emission intensity are presented. 

\begin{figure}%[htpb]
\begin{center}
\includegraphics[width=9.0 cm]{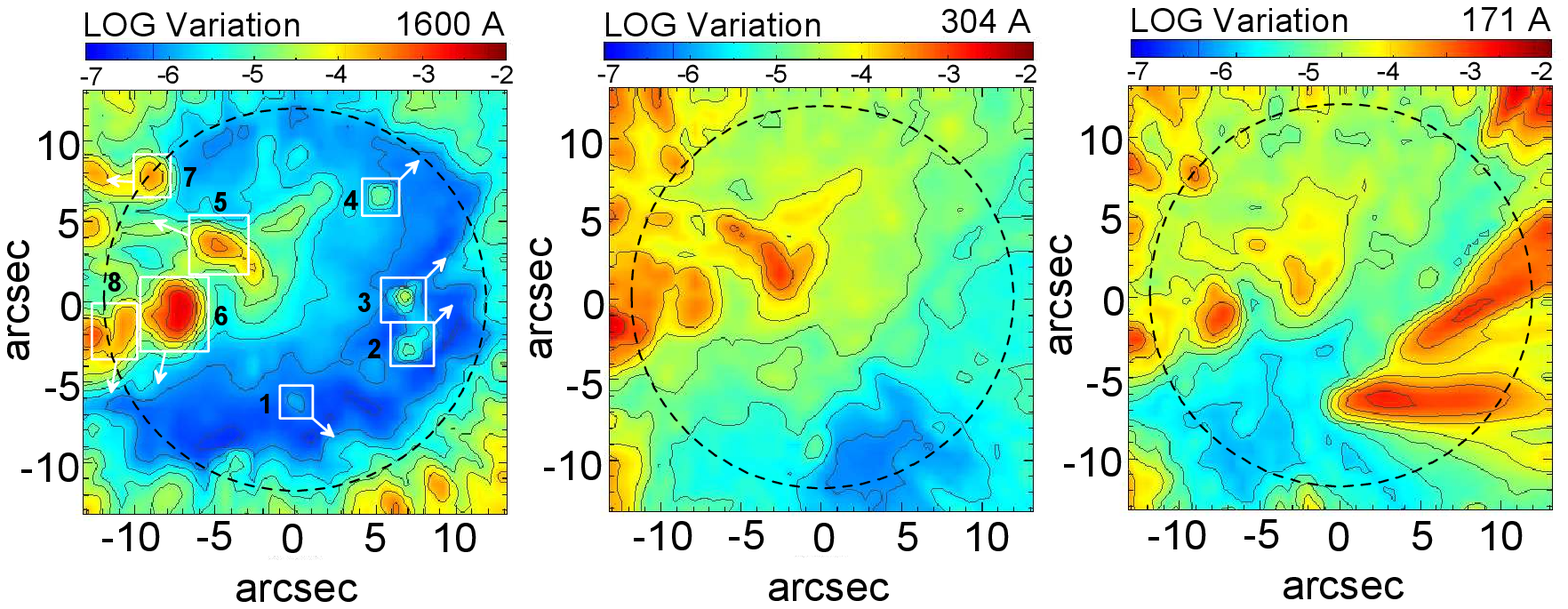}
\end{center}
\caption{Variation maps of umbral UV emission in the SDO/AIA different temperature channels (1600 \AA, 304 \AA~ and 171 \AA) obtained during 00:00-03:20 UT observation NOAA 11131 on December 08, 2010. Squares with numerals indicate the position of observed UF sources. The arrows show the scanning direction when obtaining the time-distance plots. The dash circle outlines schematically the umbral boundary. The variation intensity is presented by colours in the logarithmic scale.}
\label{2}
\end{figure}

Watching and studying frame-by-frame the films  obtained for a variety of ultraviolet wavelengths showed the presence of two dynamic components in a sunspots. The first is related to a continuous propagation of the background three-minute oscillations in the umbra and longer periodicity in the penumbra. This component is visible with a naked eye in the form of wavefronts propagating towards a penumbra from a pulsing source located in the sunspot centre. This source agrees well with the centre of the spiral wavefronts propagation described previously in \cite{2014A&A...569A..72S} for December 08, 2010 event. The other component is related to short-time brightenings of separate parts of the propagating fronts and with the emergence of small-angular size details as UF sources. 

We can see on variation maps at 1600 \AA ~(Fig.~\ref{1}, bottom panels) that the UFs sources as local brightenings have  different localizations, intensities, and shapes located in the umbral periphery. There are both bright point sources and extended sources that have different spatial orientation. Some localize near to the light bridge for example on January 10, 2016. This type of intensity variation was described in \cite{2014ApJ...792...41Y}. Watching the obtained films showed that the fast processes of the UF brightening mainly appear in the same umbral site. Also, they manifest themselves both as individual pulses and as a series of modulated pulsations. 

When we compare the obtained spatial location of bright points of variation inside umbra at 1600 \AA ~and 171 \AA, we can see well coinciding UFs sources with footpoints of coronal loops, anchored in the umbra of the sunspots (Fig.~\ref{1}, middle panels). Mainly variation maps on coronal heights show the elongated details, which can be interpret as magnetic loops along which waves propagate from the bottom layers of the sunspot atmosphere to the corona. The maxima of waves variation distributes along the loops as a bright elongated details. The main behaviour of the oscillation sources at separated periods is determined by the cut-off frequency.    

The UF source visibility varies depending on the height of the ultraviolet emission generation. We can observe a part of the flashes at all heights. The other part manifests itself only lower, at the photospheric level. The angular size of UF sources varies from flash-to-flash by revealing itself as a point or as an extended source.

\begin{figure}%[htpb]
\begin{center}
\includegraphics[width=9.0 cm]{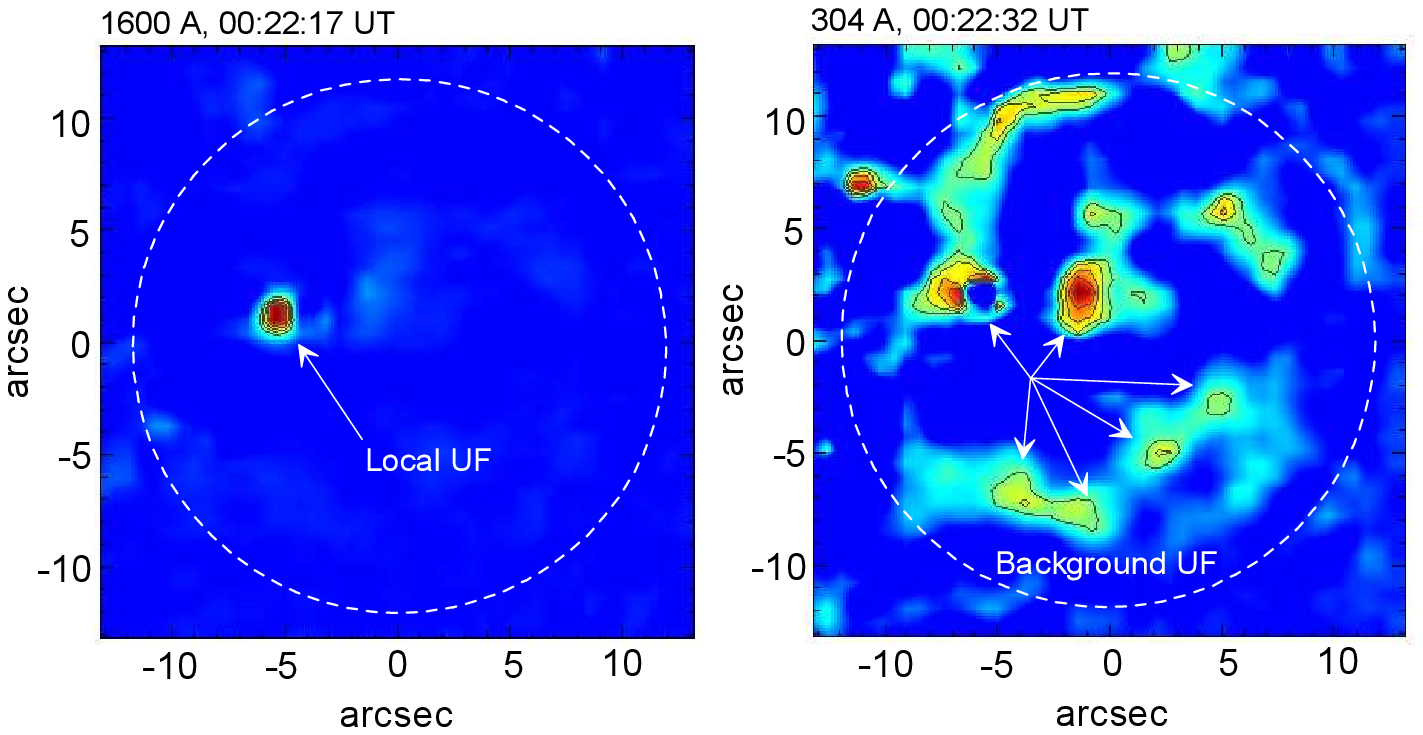}
\end{center}
\caption{Snapshots of the narrowband maps of umbral region NOAA 11131 with 3-min periodicity on December 08, 2010. The left panel shows 
the localization of the stable source of the local UFs at 1600 \AA ~(00:22:17 UT). The right panel shows the position of the bright sources at 304 \AA ~(00:22:32 UT), which ride the expending 3-min spiral wave fronts as background UFs. The dash circle outlines the umbral boundary. The arrows shows the position of UFs sources.}
\label{3}
\end{figure}

Figure~\ref{2} shows the variation maps obtained at 1600 \AA, 304 \AA, and 171 \AA ~wavelengths on December 08, 2010. One can see that the brightness variation distribution shows an inhomogeneous structure in the umbra, whose value depends on the SDO/AIA recording channel. Below, at the upper photosphere level (1600 \AA), there is a well-defined umbra indicated by the dashed circle. These umbral features have a lower level of the emission variation. Against its background, the sources that have both points and extended shapes stand out.

We found eight UF sources within the umbral boundary. The source size varies from 2 to 8 \arcsec. Mainly, these sources are located on the periphery near to the sunspot umbral boundary. When moving upwards onto the transition region level (304 \AA) we observe the disappearance of the point UF sources (No.1-4) and the increase in the brightness of the extended UF sources (No.5-8). There is an increase in the emission variation and accordingly the umbral brightness increases owing to the boost of the background three-minute oscillations. Higher, in the corona (171 \AA), we see that along with the UF sources visible below extended details appear that spatially coincide with the magnetic loops. Propagation of the background three-minute waves along these loops contributes mainly to the emission variation increase.

For the UF-type short-time processes, the maximal brightness is reached lower at the photosphere level (1600 \AA). When comparing the three-minute background component emission variations within different SDO/AIA temperature channels, the maximal value is reached at the transition region level (304 \AA). 

 The obtained variation maps show the values of the signal variance both in periodic and non-periodic components. To isolate only periodic signal, we have constructed a series of narrowband maps with 3 min signal periodicity in space and time with used the PWF technique. Figure \ref{3} shows the obtained snapshots of narrowband oscillation maps (positive half-periods) in the SDO/AIA temperature channels at 1600 \AA, 00:22:17 UT and 304 \AA, 00:22:45 UT. These times correspond to the appearance of  maximum brightness in UF source N5. We see that at wavelength 1600 \AA ~there is only one bright, local source UFs associated with periodical  oscillations in a limited spatial area. Its position almost does not change with time. At the transition zone (304 \AA), we see the wave fronts as an evolving spiral with the pulse source in the centre of umbra. Similar dynamics of wave fronts was discussed in \cite{2014A&A...569A..72S}. Contours highlight the details of the fronts, the brightness of which exceed 50 \% of the maximum value in time. With propagation waves from umbra centre to its boundary, these details continuously appear and disappear, originating the short-term brightening of separated parts of the fronts as background UFs. On the variation maps, these changes are connected with background brightening.

To understand how UF sources are related to the umbral magnetic structures, we compared their spatial position with the coronal loops seen in the UV emission (SDO/AIA, 171 \AA) and  magnetic field structure of this active region previously described in \cite{2012ApJ...756...35R}. Because the considered sunspot is the leading in the group, the magnetic field configuration shows a well-defined east-west asymmetry. The magnetic field lines anchored in the eastern part of the sunspot are much lower and more compact, than the field lines anchored in the western part of the sunspot. 

         When considering the UF source positions (Fig.~\ref{2}, 1600 \AA), we notice that the detected UF point sources (numbered 1-4) are localized in the umbral western part near to the footpoints of large magnetic loops. More extended sources (numbered 5-8) are related to the eastern part, and are located near the compact loops of the footpoints, relating the sunspot with its tail part. The size of the extended UF sources is about 7-10 \arcsec, and the point UFs are about 2.5 \arcsec.

\begin{figure}%[htpb]
\begin{center}
\includegraphics[width=9.0 cm]{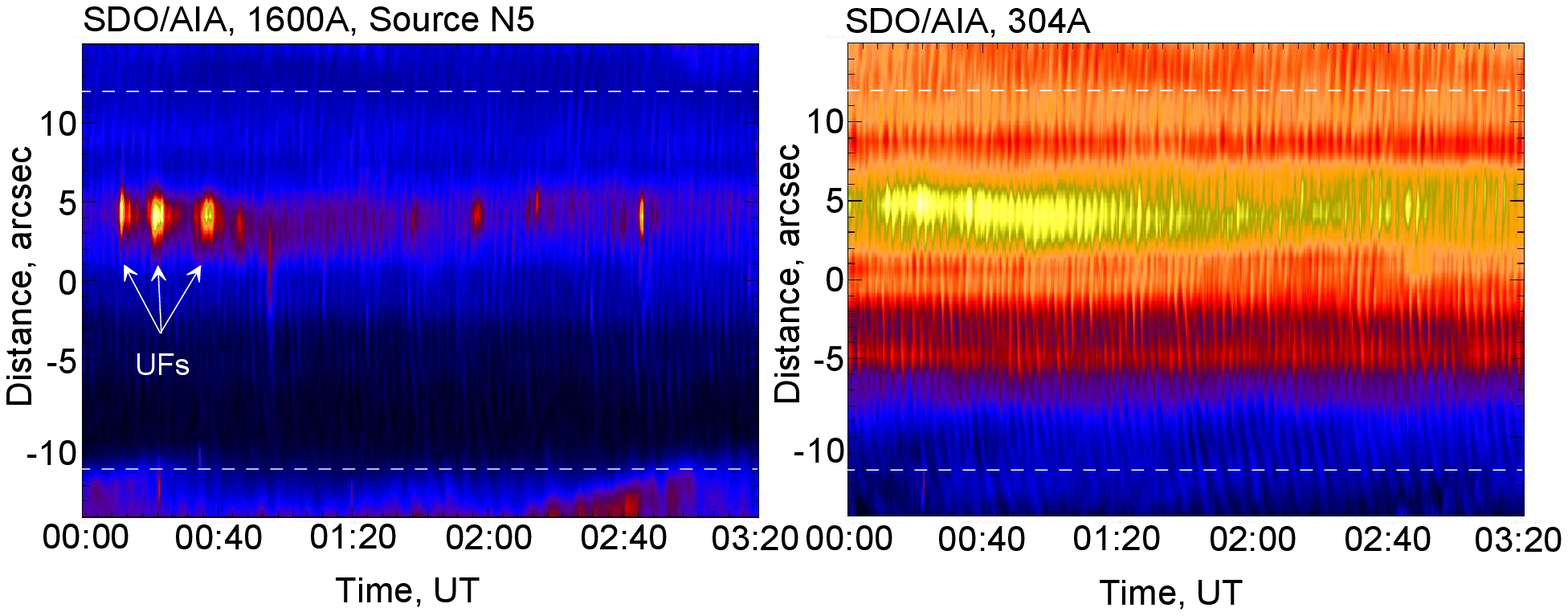}
\end{center}
\caption{Time-distance plots along the N5 UF source obtained by SDO/AIA at 1600 \AA ~(left panel), and at 304 \AA ~(right panel) temperature channels on December 08, 2010. The brightness periodic changes are the 3-minute oscillation wavefronts. The arrows show the UF. The horizontal dashed lines indicate the umbra/penumbra border. The 1D spatial coordinates are in arcsec, the time in UT.}
\label{4}
\end{figure}

\subsection{Time dynamics of UFs on December 08, 2010}

More comprehensive analysis of the time dynamics for wave processes was performed for the sunspot active region NOAA 11131 on December 08, 2010. The wave processes inside the umbra were intensively studied by \cite{2012A&A...539A..23S, 2014A&A...569A..72S, 2014A&A...561A..19Y, 2014AstL...40..576Z}.

The detected compact sources of the maximal variation in Fig.~\ref{2} were studied to reveal the existence of flash and/or oscillation activity. For this we scanned each of the sources at 1600 \AA ~and 304 \AA and built the time-distance plots. The arrows  show the UF source scan directions.

\begin{figure}%[htpb]
\begin{center}
\includegraphics[width=9.0 cm]{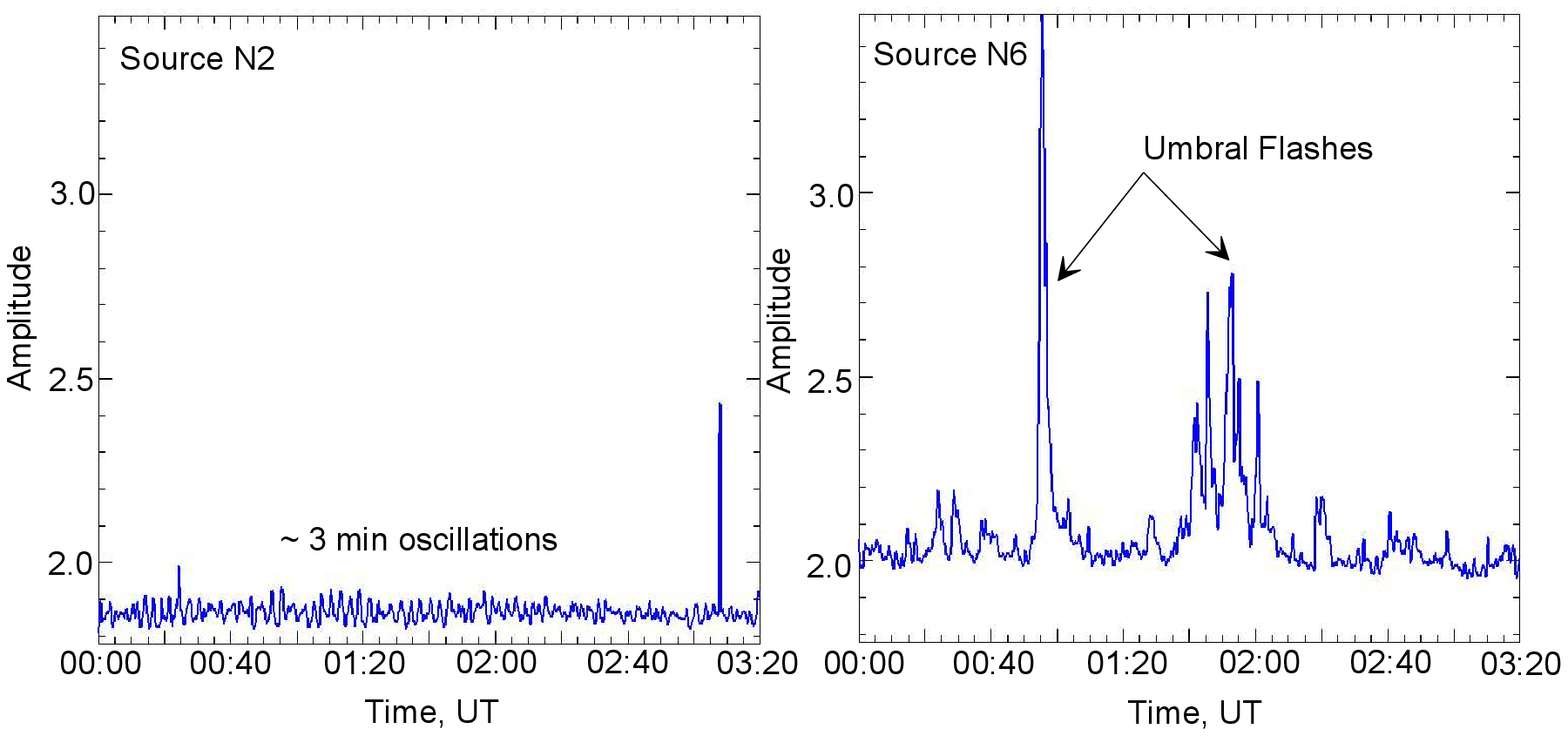}
\end{center}
\caption{Time dynamics of the EUV emission for the N2 and N6 sources at 1600 \AA. The arrows show the maximum emission of UF. Time in UT.}
\label{5}
\end{figure}

Figure~\ref{4} presents an example of the obtained time-distance plots in 1600 \AA ~(left panel) and in 304 \AA ~(right panel) for the N5 extended source. We see that throughout the entire observational time there are background three-minute broad brightness variations in the umbra that smoothly transit into the five-minute oscillations at the boundary of the umbra  and penumbra shown by the dashed line. This type of partial brightening of wave fronts during propagation in the umbra as UFs was described in \cite{2014ApJ...792...41Y}.  Most clearly, these UFs are exhibited at the level of the transition region in 304 \AA ~(Fig.~\ref{4}, right panel). Also, these oscillations exist lower, at the level of the upper photosphere (1600 \AA). Against their background, we note a series of periodically recurrent various-power local UFs. The arrows in Fig.~\ref{4} indicate separate pulses (left panel). The position of flashes by space coincides with the maximal brightness of the N5 extended source. The fine spatio-temporal structure of the UF sources also coincides with the brightenings of the three-minute oscillation background wavefronts.

When comparing the flash peak values below and above the sunspot atmosphere, we note that UFs have shorter duration at the level of the photosphere than that  at the level of the transition region. Low-frequency modulation of three-minute oscillations occurs. The brightness change at 304 \AA ~occurs smoothly without well-defined peaks. During flashes brightenings of the 3-minute wavefront in the source occur. The brightness contrast decreases as the height of the UF observation increases. One may assume that UFs and the background three-minute oscillations have identical natures in the form of the wave activity increase within the magnetic loops, where their propagation occurs with different time and spatial scales.

To compare the time profiles of the brightness variation within different UF sources for one wavelength, we used cuts along spatial coordinates with the maximal brightness on the time-distance plots (Fig.~\ref{4}). The profiles for each UF source were obtained. Fig.~\ref{5} shows a brightness change example for N2 and N6 sources at the level of the upper photosphere (1600 \AA), where the UF visibility is maximal.

One can see that, along with the well-defined three-minute oscillations (Fig.~\ref{5}, left panel), there also exist pulse events as UFs. Their number and duration depends on the flash source. Thus, we only observed individual flashes during a three-hour interval of observations for the sources numbered 1 through 4. At the same time, on the profiles of the 5-8 sources, we note series of flashes with different amplitudes and durations (Fig.~\ref{5}, right panel).

Comparing the shape of the revealed sources in Fig.~\ref{2} with the corresponding profiles in Fig.~\ref{5} showed that, for point sources, the emergence of rare individual UFs is common. The UF extended sources are related to the series of periodically recurring different amplitude pulses, about 4-14 flashes during the observations. Comparing the peak amplitudes of various UF sources revealed that the brightness change in the point sources is almost five times less, than that for the extended.

\begin{figure}%[htpb]
\begin{center}
\includegraphics[width=9.0 cm]{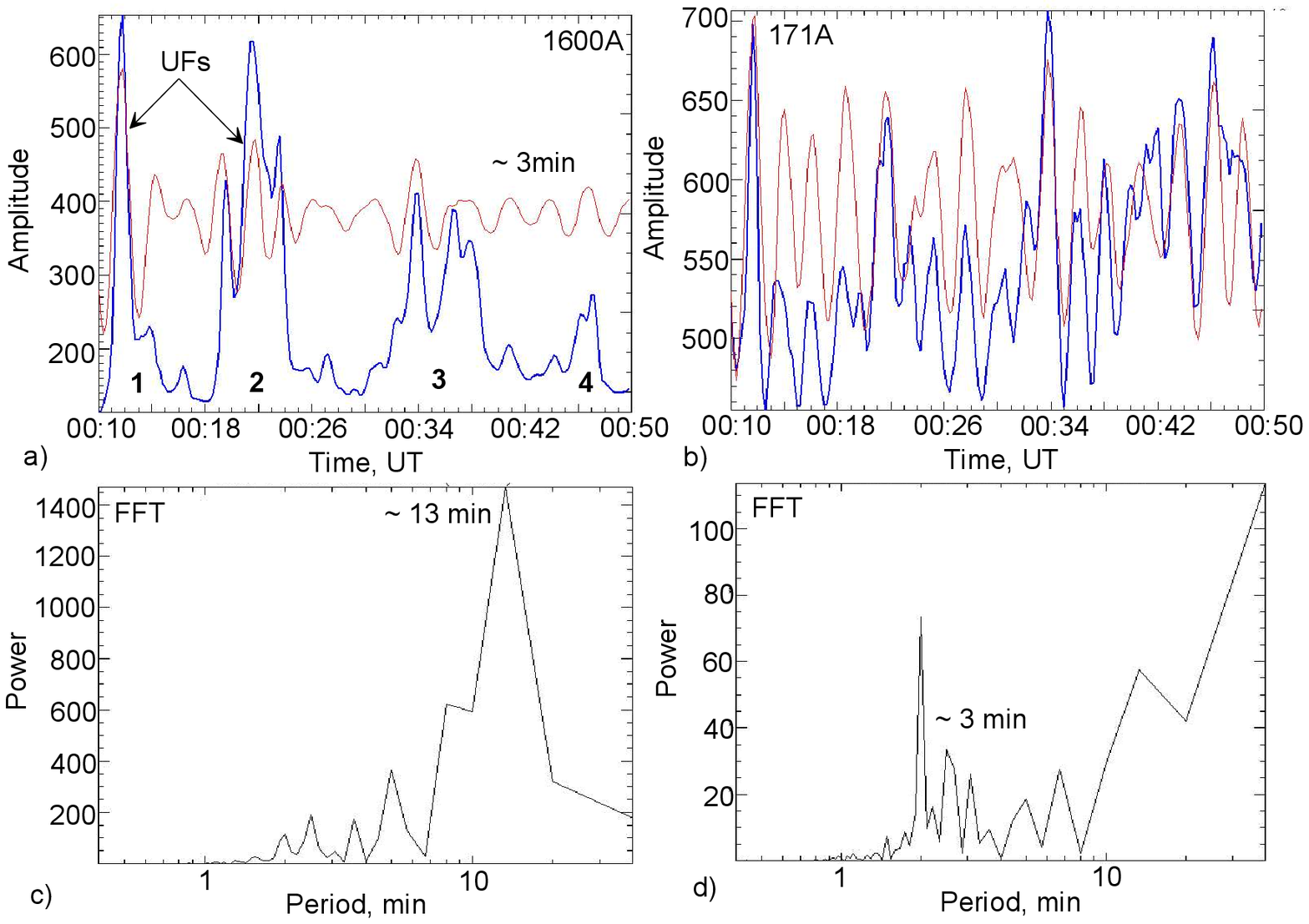}
\end{center}
\caption{Time dynamics for the N5 UF source at various SDO/AIA channels: 1600 \AA, left panel and 171 \AA, right panel. The blue lines show the brightness changes recorded during the flashes. The red lines show the time profiles of the filtered 3-minute oscillations. The numerals denote the oscillation train numbers. Bottom panels: Fourier spectra of UF signals for SDO/AIA channels accordingly.}
\label{6}
\end{figure}

\subsubsection{Relation between wave dynamics and UFs}

Based on the obtained 1D time-distance plots for each source ~(Fig.~\ref{4}), for which the relation between the oscillating 3-minute component and the UF emergence is well traced, we performed a spectral analysis of the time profiles by using the fast Fourier transform (FFT), and PWF technique. We applied the Fourier transform to provide a good spectral resolution, and the PWF technique to obtain a spatio-temporal structure of the wavefronts propagating in UF sources.

Figure~\ref{6} shows an example of the oscillations detected in the N5 extended source over the 00:10-00:50 UT observational period, when there emerged UFs. We can see the profiles with sharp UFs at 1600 \AA. At the corona level at 171 \AA ~there are stable 3 min oscillations without spikes. This served as the main criterion for studying the spectral behaviour of filtered 3 min oscillations at 171 \AA ~and its comparison with the original signal at 1600 \AA. In this case the spectral power does not change because of sharp jumps in the signals.

One can see that at the level of the upper photosphere (Fig.~\ref{6}a, 1600 \AA, blue lines), there exist periodic brightness changes in the EUV emission. These changes take shape as a UF series, where UFs were exhibited as a sequence of low-frequency trains of higher frequency oscillations. Those higher frequency oscillations are particularly expressed in the sunspot atmosphere higher coronal layers at 171 \AA ~(Fig.~\ref{6}b). The Fourier spectrum showed the existence of significant harmonics. These harmonics are related to an $\sim$ 3-5-minute periodicity and to the $\sim$ 13-min low-frequency oscillations (Fig.~\ref{6}c,d).

To trace the time dynamics for the detected periodicity, we performed a wavelet filtration of the series in the period band near three minutes. We found the four trains of high-frequency oscillations numbered in Fig.~\ref{6}a. If one compares the behaviour of the filtered three-minute signal (red lines) and the UF emergence (blue lines), it is apparent that the train maxima coincide with the UF brightness maxima. A complex UF time profile (in the form of a series of variable peaks) is related to the existence of oscillations with different amplitudes, phases, and lifetimes in the trains.

When comparing the oscillations in UFs, one can see (Fig.~\ref{6}), that the low-frequency trains are well visible in the lower atmosphere. Their power decreases in the upper atmosphere. This is well traced on the Fourier spectra of the signals for different height levels (Fig.~\ref{6}c,d). We note the inverse dependence between the harmonic power. At the level of the upper photosphere, the low-frequency modulation is maximal at a low level of the 3-minute harmonic. In contrast, in the corona, there is a pronounced peak of 3-minute oscillations with the minimal value of the $\sim$ 13-minute component power.

Increasing oscillations in the source led to the formation of compact brightenings in the form of UFs on the time-distance plot (Fig.~\ref{4}, left panel). As the low-frequency oscillation power decreases, at the corona level a smooth increase occurs in the high-frequency three-minute component in the form of brightenings of the wavefront separate details (Fig.~\ref{4}, right panel). The mean UF duration for extended sources was $\sim$ 3.7 minutes. This value is near the value of one period for the three-minute oscillation maximal power.

To test the obtained association between UFs and oscillations, we calculated the correlation coefficients between the original signal and the three-minute filtered signal in various SDO/AIA channels. There is a direct correlation between the three-minute oscillation power and the UFs  power. The maximal value for the correlation coefficient is at 1600 \AA, and this value varies within the 0.65 - 0.85 range for different sources of flashes.

One may assume that the obtained association between the increase in the three-minute oscillations and the UF emergence is characteristic of not only the detected N5 source, but is also present in all the detected sources. To test this statement, we calculated the narrowband three-minute oscillation power variations in the N7 and N8 sources above, at the corona level (171 \AA), and compared these variations with the UF emergence in the integral signal lower, at the photosphere level (1600 \AA). The observational interval was 00:00-03:20 UT.

\begin{figure}%[htpb]
\begin{center}
\includegraphics[width=9.0 cm]{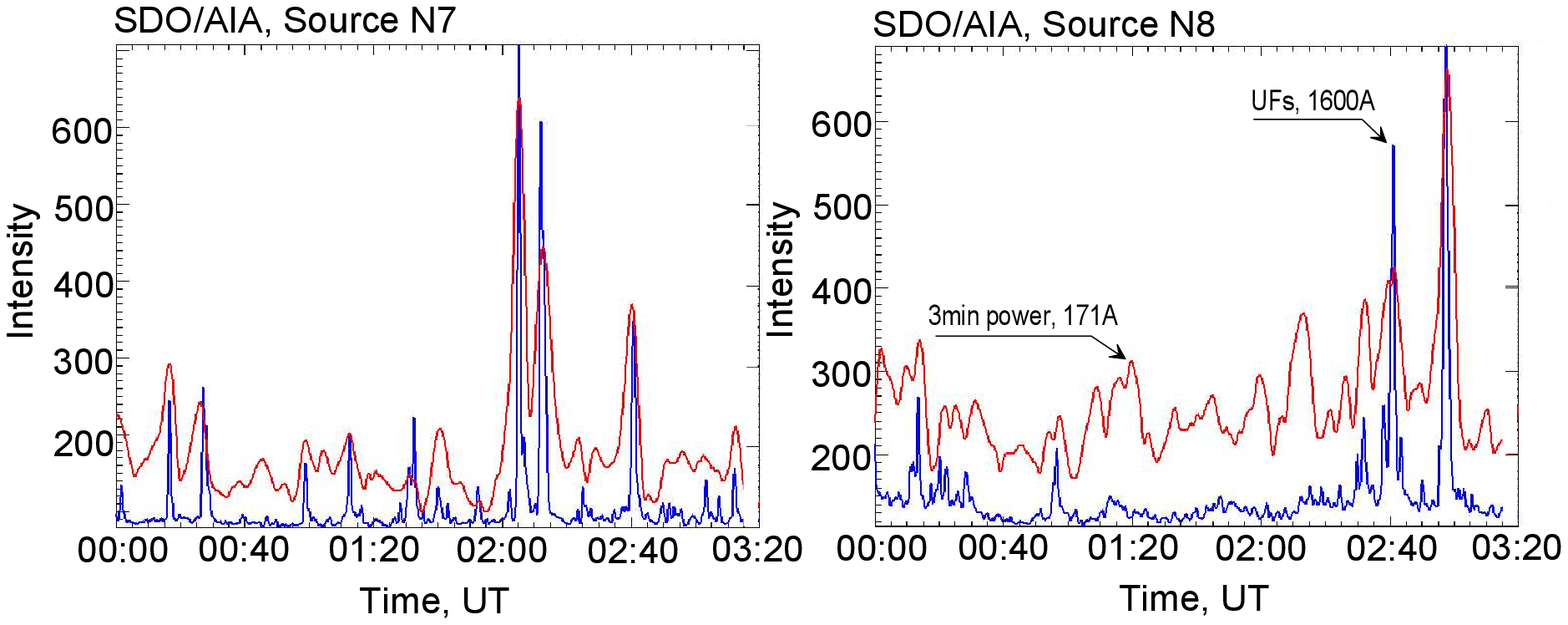}
\end{center}
\caption{Amplitude variations of the N7 and N8 extended sources of UFs at 1600 \AA ~and 171 \AA ~temperature channels. Blue lines show the profiles of the original signal at 1600 \AA. Red lines show the 3-min oscillation power at 171 \AA.}
\label{7}
\end{figure}

Figure~\ref{7}  shows the time profiles for the signals in the N7 and N8 extended sources, and the corresponding variation of power oscillations in the corona. Apparently, in the sources at the upper photosphere level (blue lines, 1600 \AA), there are recurrent UFs of different amplitude. In addition to  the case with the N5 source, the bulk of the UF peak values are accompanied by an increase in the three-minute oscillation low-frequency trains at the corona level (red lines, 171 \AA). There is a well-defined correlation between the signals. Thus, over 01:20-03:20 UT, the emergence of the "step-like" signals at the photosphere level with their gradual steeping and the emergence of UF pulses is followed by a smoothly varying increase in the power of the three-minute oscillation trains in the corona.

\begin{figure}%[htpb]
\begin{center}
\includegraphics[width=9.0 cm]{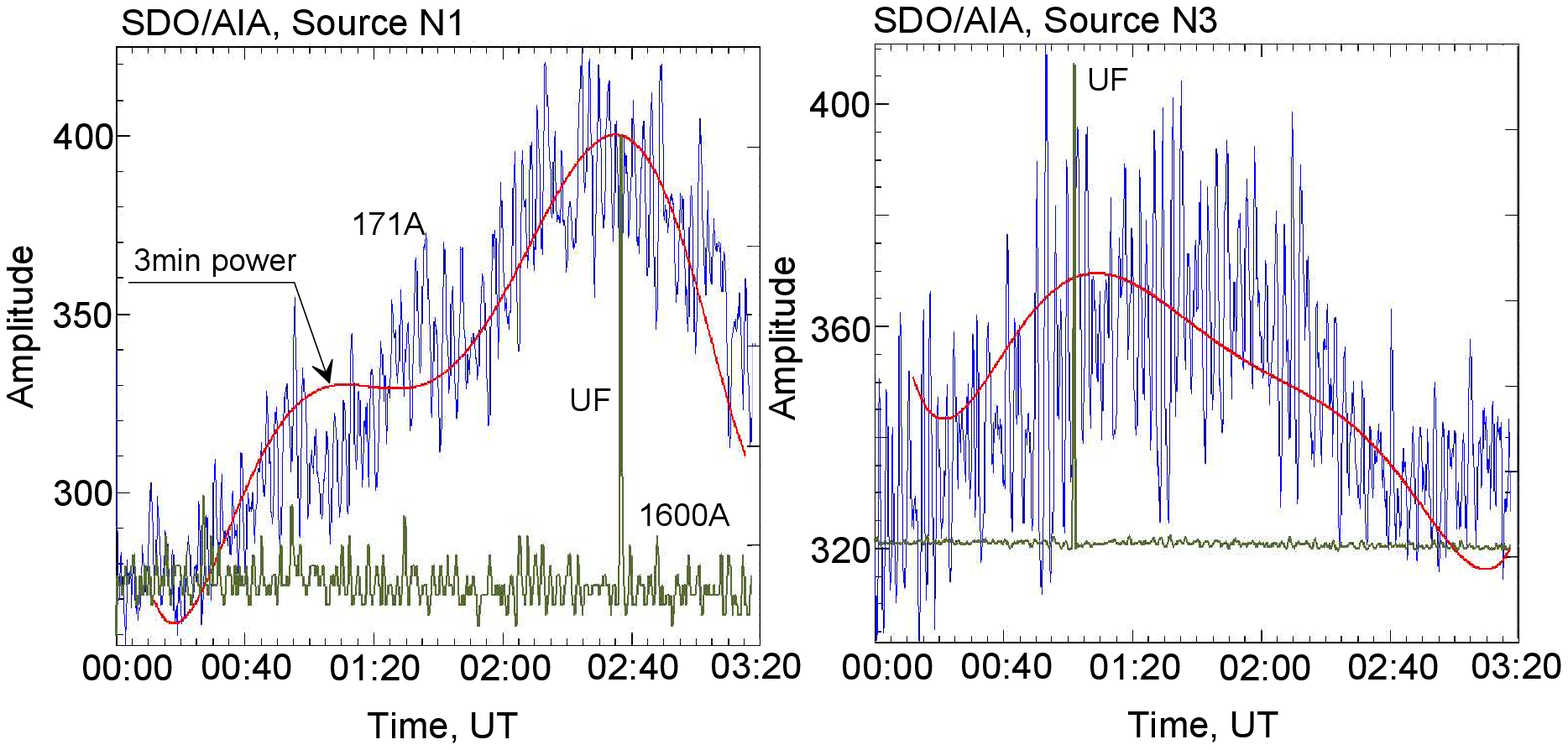}
\end{center}
\caption{Amplitude variations of the point N1 and N3 UF sources. Green lines show the original signal at 1600 \AA; blue lines present the signal at 171 \AA. Red lines show the mean power of the 3-min oscillations.}
\label{8}
\end{figure}

\begin{figure*}%[htpb]
\begin{center}
\includegraphics[width=14.0 cm]{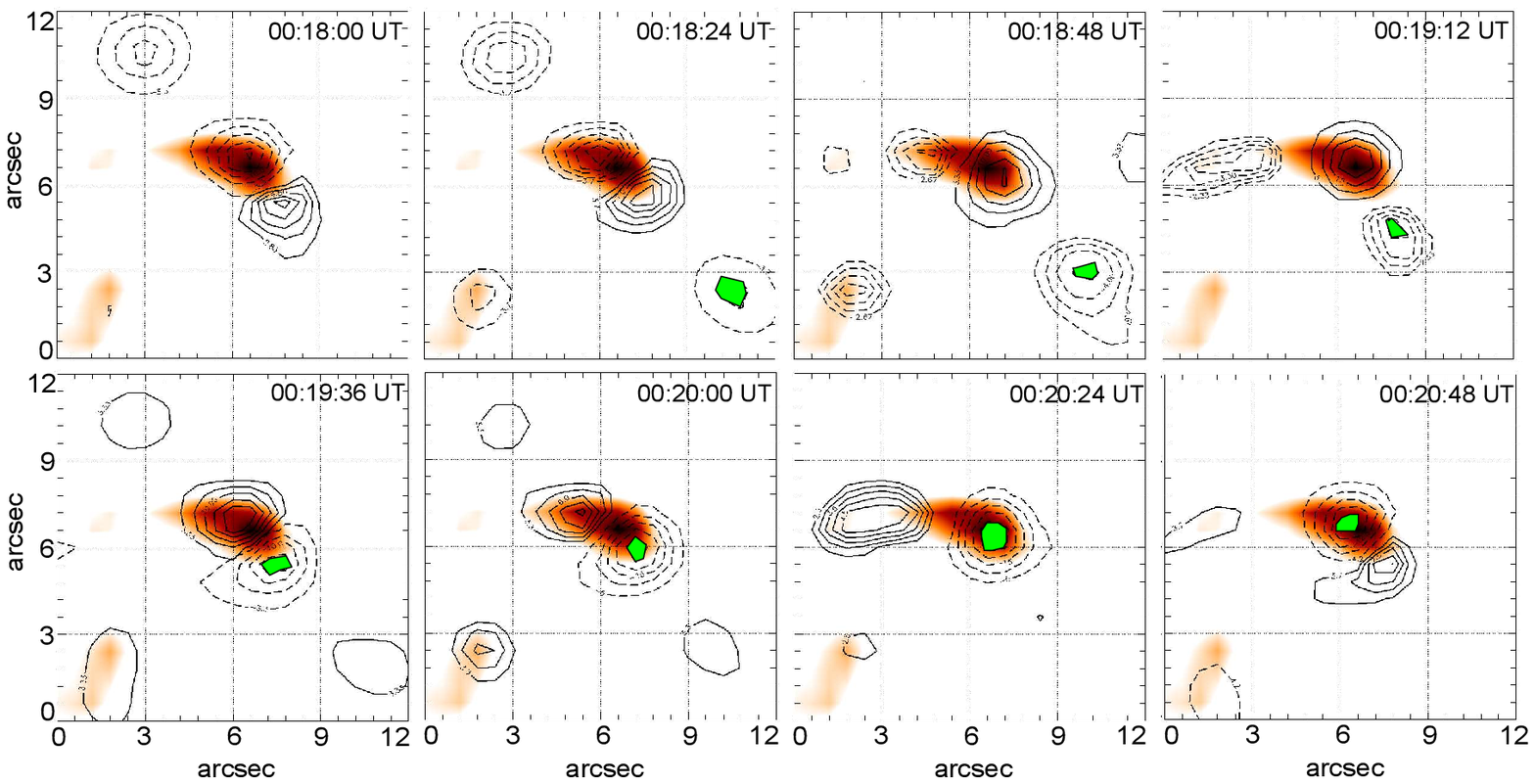}
\end{center}
\caption{Snapshots of spatial distribution of travelling wave fronts during the UF for the N5 extended source. Duration of propagating the 3-min waves along a magnetic waveguide of about one period. The observational wavelength is 1600 \AA. A continuous line represents a positive half-period of propagating waves, and the dashed line separates the negative half-period. The background image is the distribution of the brightness of the source at the time of the maximum flash. The minimum negative half-period is indicated in green. The time resolution is 24 sec.}
\label{9}
\end{figure*}

For the N1 and N4 point sources only single pulses with a low-intensity level were observed. For these sources, we compared the coronal three-minute oscillation power mean-level variations with the moments of the UF single burst peak emergence at the photosphere level. Fig.~\ref{8} shows the original signal profiles at varying height levels (green lines for 1600 \AA, blue lines for 171 \AA) with the superposition of the three-minute oscillation mean power (red lines, 171 \AA). Apparently, the moments of the short flash emergence below the sunspot coincide with the three-minute oscillation power maxima above. In this, we note a similar sequence in the signal evolution such as that for the extended sources. The difference is in the duration of the flashes. Thus, for the N1 source (02:36:15 UT), the UF duration was $\sim$ 1.5 minutes, for N2 (03:07:30 UT) - about 1.1 minutes, for N3 (01:01:30 UT) - about 1.0 minute, and for N4 (03:12:00 UT) - about 1.1 minutes. The UF mean value for the point sources was $\sim$ 1.2 minutes.

\subsubsection{Wave propagation in UF sources}

To study narrowband wave propagation over the UF source space, we used the PWF technique. Fig.~\ref{9} shows the time sequence for the EUV emission wavefront images (SDO/AIA, 1600 \AA) obtained for the N5 source during the second train of the three-minute oscillations (00:18:00 - 00:20:48 UT). The temporal resolution was 24 sec. The oscillation positive half-period is shown by the continuous contours, the negative is outlined by the dash contours. The basis is the source image at the UF maximum instant at 00:20 UT.

Comparing the obtained images (Fig.~\ref{9}) with the profile for the UF maximal brightness variation (Fig.~\ref{6}a), we can clearly see that the brightness increase is accompanied by the onset of the wave propagation along the detected direction coinciding (by shape) with the UF maximal emission source. These motion directions towards the penumbra, in turn, coincide with the footpoint of the magnetic loop, along which the waves propagate. There are recurrent instances when the fronts emerge in the same site of the umbra limited space. The beginning of the N5 extended source coincides (by space) with the pulsing centre of the three-minute waves expanding spirally. One may assume that the wave source coincides with the footpoint of the magnetic bundle that diverges in the upper atmosphere. Separate spiral arms rotate anti-clockwise. These background waves were studied in \cite{2014A&A...569A..72S} for this active region.

 Presumably, propagation of spiral-shaped waves (Fig.~\ref{3}, 304 \AA) is the initiator of the wave increase in separate magnetic loops. In this case, the bulk of bright fronts propagates towards the bright extended UF emergences. The wave propagation projection velocities along the waveguide lie within the 20-30 km/s interval. These values agree with the slow magneto-acoustic wave propagation velocity in the sunspot.

For different low-frequency numbered trains of the UF in the N5 source (Fig.~\ref{6}a, 1600 \AA), the maximal brightness was located in various parts of the magnetic waveguide, and it varied with time. Each series UFs with the $\sim$  10-13 minute duration was accompanied by an increase in the low-frequency trains of the 3-minute waves. There are differences for each wave train. One observes the UFs, when both propagating and standing waves are visible throughout one train. The wave velocity can vary from train to train. Mainly, the waves move towards the sunspot boundary from its centre.

The increase in the wave processes for the UF point sources occurs in the form of producing single pulses in the umbral site limited to several pixels. The emergence of so-called standing waves without their apparent propagation is characteristic for these sources. Mainly, the 2D time dynamics of the three-minute oscillation sources agrees with the UF source dynamics.

\section{Discussion}

The results obtained from the SDO/AIA data showed that the investigated phenomenon of UFs is characteristic of all the heights within a sunspot atmosphere. We see a response to  both below, at the photosphere level, and above, at the corona level, the sunspot atmosphere. This means that flashes represent a global process of energy release and this process encompasses all the layers of an sunspot umbra.
 
Usually, an umbra is considered a relatively quiet region as compared with a penumbra. This is because the umbral magnetic field represents a vertical bundle of magnetic field lines diverging with height. The umbral field line inclination is minimal. Correspondingly, the magnetic reconnection responsible for the flash energy release emergence is unlikely in a homogeneous, vertical field. This conclusion indicates that there are other mechanisms for the emission increase during UFs.

A wave mechanism is an alternative to explain this increase. It is based on the assumption that the observed brightenings in the form of UFs are short-time power increase in wave processes within separate umbral parts. This viewpoint to be common, because the well-known three-minute umbral oscillation were revealed to propagate non-uniformly both over the sunspot space and with time \citep{2012A&A...539A..23S, 2014A&A...569A..72S}. Mainly, the waves are modulated by a low-frequency component in the form of the $\sim$ 13-15 minute trains and their power is time variable. The wave motion direction is determined by the spatial structure of the umbra-anchored magnetic field lines, along which slow magneto-acoustic waves propagate.

There are instances when a significant increase in power of the three-minute oscillation trains occurs at separate footpoints of magnetic loops. These processes have an indefinite character, and the source of the next wave increase is impossible to locate. On the other hand, the magnetic loop footpoints are stable over the umbral space over a certain time period. This enables us to assume that the positions of the UF sources are probably directly related to the magnetic loop footpoints, in which short-time increases in the three-minute waves are observed.

        These assumptions agree well with the spatial localization of the UF sources at the umbral boundary (Fig.~\ref{1}) as well as the difference in shape, i.e. extended and point. Umbral flash sources maintain their spatial stability for about three hours, producing UF series. On the other hand, \cite{2003A&A...403..277R} noted that some flashes possess instability both in space and time. 
        
In \cite{2014ApJ...792...41Y}, the authors showed that the UFs visible on time-distance plots occur at random locations without a well-established occurrence rate. It has been established that the appearance of new UFs sources is associated with the trains of three-minute oscillations in the sunspot umbra with much larger amplitude. The individual UFs ride wave fronts of umbral oscillations. A possible explanation for this is the presence in the umbra background oscillations as expending fronts of 3-min waves and their interaction between each other. A similar type of brightening was considered in \cite{2014A&A...569A..72S}. These authors noted that the individual parts of the wave fronts, which are shaped as rings or spirals, during propagation along magnetic loops with different
spatial configuration and interactions between each other, can lead to the appearance of  diffuse brightening  
with spatial instability. Such short-lived background UFs are well visible on the time-distance diagrams, constantly appear in umbra, and do not have stable shapes and localizations in space (Fig.~\ref{3}, 304 \AA). Basically, the pulse source of such wave fronts is located in the centre of umbra, and is possibly associated with the footpoint of the magnetic bundle whose loops are expanding with height. 

In the case of background UFs, we observed the local traces of waves that propagate along loops with different inclinations relative to the solar normal and correspondingly  different cut-off frequencies. This forms a brightening of wave tracks, which we observed as diffuse UFs during increasing of power oscillations in selected areas of umbra. We can also obtain the same effect during interactions between wave fronts. With height, the visibility and positions in space of these sources are shifted in a radial direction because of upwards wave  propagation.

For the local UFs discussed in our work, the sources have small angular size with a periodic 3-min component and stable location, both over space and height (Fig.~\ref{3}, 1600 \AA). Their appearance is associated with the power of the maximum wave propagating near the footpoints of coronal loops outside the main magnetic bundle. The origin of these loops is umbral periphery. Their inclination can be different relatively to the configuration of the main magnetic bundle.

        The existence of an UFs fine structure was previously assumed in \cite{2000Sci...288.1396S} and \cite{2005ApJ...635..670C} using spectroscopic observations. Improving the angular and spatial resolutions of astronomical instruments enabled us to observe such changes in UF sources directly. Thus, \cite{2009ApJ...696.1683S} used HINODE data (CaII H line) to find an umbra fine structure in the form of a filamentary structures that emerged during UFs. These details were present at an oscillation increase, and formed a system of extended filaments, along which brightness varied with time in the form of travelling bright and dark details. The calculated horizontal velocity varied within 30-100 km/s.
        
We can assume that we observe in UF sources projection motions (at the footpoints of magnetic field lines) of the three-minute longitudinal wavefronts propagating upwards \citep{2003ApJ...599..626B}. Depending on the field line start localization (the umbral centre or near its boundary) and on the inclination to the solar normal, there is a projection effect of wave visibility. Near the sunspot boundary, one observes extended UF sources, whereas closer to the centre point sources are observable. This statement is true if we assume a symmetry of diverging field lines relative to the umbral central part. In reality, there is often an east-west asymmetry of an active group. This asymmetry is related to the presence of the head (sunspot) and tail (floccule) parts.

The wave path length, and, accordingly, the wave visibility with certain oscillation periods is determined by the cut-off frequency \citep{1977A&A....55..239B, 1984A&A...133..333Z}. The path also varies as a cosine of the magnetic waveguide inclination angle. Point UF sources with a minimal angular size are related to the footpoints of vertical magnetic field lines. Large, extended UF sources, are related to the footpoints of field lines with a large inclination to the solar normal.

Comparing the positions of the sources at the NOAA 11131, various heights showed a good concurrence of the UF sources underneath (1600 \AA) with the footpoints of coronal loops (171 \AA), which play the role of magnetic waveguides for three-minute waves. For the low-lying loops in the eastern part of the NOAA 11131 that connect the sunspot with the tail part, we see extended UF sources at their footpoints. For the western part, we see point sources.

The revealed interconnection between the UF emergence and the increase in the three-minute wave trains indicates that we can consider UFs as events, in which peak increases in the trains of oscillations at the footpoints of magnetic loops exhibit themselves. There is a direct dependence between the oscillation power and the flash brightness at a maximal correlation. The higher the magnitude of the three-minute waves, the more powerful the flash. This dependence concerns both extended UF sources and point UF\ sources. The UF emission maximum coincides with the maximum of the three-minute oscillations within one wave train.

The 2D spectral PWF-analysis for the SDO/AIA image cube directly showed (Fig.~\ref{9}) that, during UFs, three-minute wave motions emerge along the detected magnetic loops towards the umbral boundary at the UF sources. The wave propagation starts at the footpoint and terminates at the level, where its inclination angle corresponds to the cut-off frequency beyond the limits of the observed three-minute frequency band. The greater the loop inclination, the greater the projection of the UF source to the observer, and the more wave trains (UF pulses) we can record. Correspondingly, we will observe extended, bright UF sources. Contrary to the propagating waves, so-called standing waves will be observed for point sources. An explanation of this is the projection of the propagating waves along vertical magnetic loops towards the observer. In this case,  spatial wavefronts will be observed within a loop cut region limited by space. Those fronts form UF sources with a small angular size.

The UF source lifetime will also be different. For point UFs, the source lifetime is about 1-2 minutes; for extended UFs, it is 3-15 minutes. The visibility of the source is restricted by a low integral power level of the UF emission of the point sources and a short observational time for maximal oscillations (1-2 half-periods). For extended UF sources, we can observe a few travelling wave trains simultaneously, which intensifies their integral brightness and increases the observational time (lifetime).

\section{Conclusions}

We analysed the association between an increase of wave activity in the sunspot active groups and the emergence of UFs. We used the observational data in the UV emission obtained in the various temperature channels of the SDO/AIA with high spatial and temporal resolutions. To detect the oscillations, we used the time-distance plots and Fourier and wavelet transform spectral techniques. The results are as follows:

1)      We revealed fast periodic disturbances related to the wave activity in the sunspot umbra during a three-hour observation. These disturbances correlate well with the continuous diffuse brightening of separate details of the propagating three-minute wavefronts as described in \cite{2014ApJ...792...41Y}. Along with this, short-time emergences of the small local sources having a periodic component and identified as UFs are observed.

2)      We can divide the observed umbral brightening into two types. The first type are background UFs associated with random brightening of separated parts of wave fronts during their propagation. These UFs are observed all of the time in umbra as weak diffuse details that ride the wave fronts without stable shapes and localization in space. The second type are local UFs associated with the increasing of wave activity near to the  footpoints of magnetic loops. These sources not change spatial position in time and show pronounced wave dynamics during UFs.

3). For the local UFs we revealed various types of spatial shapes of the sources. We suppose that the point sources are located at the footpoints of large magnetic loops. Their feature is the activity with rare single pulses of low power and duration. The extended sources are related to the footpoints of low magnetic loops and large inclinations. The features of this source type are series of recurrent UF pulses related to propagating trains of three-minute waves. The flash power depends on the distance of the wave path, along which the emission is integrated. The wave path and, correspondingly the UF source projection size, are determined by the cut-off frequency.

4)      The emergence of the main UF maximum is shown to coincide with the maximal value of the power of the three-minute oscillation trains in separated loops. This type of wave dynamics follows that described in \cite{2014ApJ...792...41Y} for background UFs but localized in magnetic loops. There is a correlation between the UF emergence at the photosphere level and the increase in the power of the three-minute wave trains in the corona. 

These results explicitly show the correlation between the sunspot three-minute oscillation processes and the UF emergence. These processes are a reflection of the slow magneto-acoustic wave propagation from the subphotospheric level into the corona along the inclined magnetic fields. The wave process power dynamics in separate magnetic loops determines the time and site of the UF source emergence. The main mechanism responsible for the observed UF parameters is the wave cut-off frequency. In the future we plan to study in more detail the relationship between the shape of the local UFs sources and the inclination of the magnetic loops near to the footpoints of which the flashes are observed.

\begin{acknowledgements}
We are grateful to the referee for helpful and constructive comments and suggestions. The authors are grateful to the SDO/AIA teams for operating the instruments and performing the basic data reduction, and especially, for the open data policy. This work is partially supported by the Ministry of Education and Science of the Russian Federation, the Siberian Branch of the Russian Academy of Sciences
(Project II.16.3.2) and by the programme of basic research of the RAS Presidium No.28. The work is carried out as part of Goszadanie 2018, project No. 007-00163-18-00 of 12.01.2018 and supported by the Russian Foundation for Basic Research (RFBR), grants Nos. 14-0291157 and 
17-52-80064 BRICS-a. The research was funded by Chinese Academy of Sciences President’s International Fellowship Initiative, Grant     No. 2015VMA014. 
\end{acknowledgements}

\bibliographystyle{aa} % style aa.bst
\bibliography{BibTeX_Sych} % your references Yourfile.bib

\end{document}